\documentclass[prl,secnumarabic,amssymb, nobibnotes, aps, prd,aps,twocolumn]{revtex4-1}
\usepackage{graphicx , epsfig,color}
\usepackage{amsmath}
\usepackage{latexsym}
\usepackage{amstext}
\usepackage{array}
\usepackage{multirow}
\usepackage{changebar}
\usepackage{subfigure}
\usepackage{stackrel}
\usepackage{ulem}
\usepackage{mathtools}

\newcommand{\mb}[1]{ \mbox{\boldmath$#1$} }
\newcommand{\ds}{\displaystyle}
\newcommand{\beq}{\begin{eqnarray}}
\newcommand{\eeq}{\end{eqnarray}}
\newcommand{\beqq}{\begin{eqnarray*}}
\newcommand{\eeqq}{\end{eqnarray*}}

\newcommand{\eps}{\varepsilon}

\newcommand{\x}{\mbox{\boldmath$x$}}

\begin{document}
\title{ \textbf{Hybrid Markov-mass action law for cell activation by rare binding events}}
\author{C. Guerrier $^{1}$ and D. Holcman$^{1,2}$} \affiliation{$^1$ Computational Biology and Applied Mathematics, Ecole Normale Sup\'erieure Paris, France.  $^2$ DAMTP and Churchill College CB3 0DS, Cambridge United Kingdom. \footnote{ Mathematical Institute, Oxford OX2 6GG, Newton Institute. This research is supported by a Marie Curie Award.}}
\date{\today}

\begin{abstract}
 The binding of molecules, ions or proteins to specific target sites is a generic step for cell activation. However, this step relies on rare events where stochastic particles located in a large bulk are searching for small and often hidden targets and thus remains difficult to study. We present here a hybrid discrete-continuum model where the large ensemble of particles is described by mass-action laws. The rare discrete binding events are modeled by a Markov chain for the encounter of a finite number of small targets by few Brownian particles, for which the arrival time is Poissonian. This model is applied for predicting the time distribution of vesicular release at neuronal synapses that remains elusive. This release is triggered by the binding of few calcium ions that can originate either from the synaptic bulk or from the transient entry through calcium channels. We report that the distribution of release time is bimodal although triggered by a single fast action potential: while the first peak follows a stimulation, the second corresponds to the random arrival over much longer time of ions located in the bulk to small binding targets. To conclude, the present multiscale stochastic chemical reaction modeling allows studying cellular events based on integrating discrete molecular events over various time scales.
\end{abstract}
\maketitle
\begin{figure}[http!]
 \centering
 \includegraphics[scale=0.45]{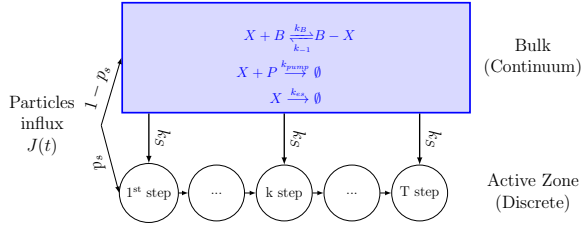}
\caption{{\bf Coarse-grained Markov-mass action model:} the mass action equations in the bulk are coupled to the Markov chain that models the arrival of single particles to the target sites until the threshold $T$ is reached.}
\label{MarkovModel}
\end{figure}
\begin{figure}[http!]
  \centering
 \includegraphics[scale=0.5]{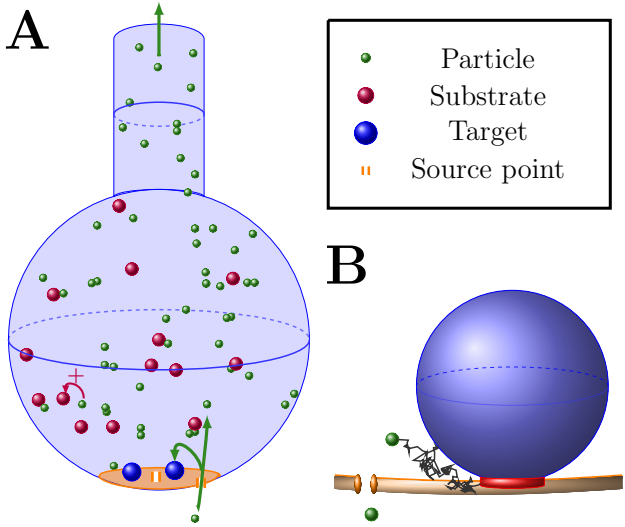}
\caption{ {\bf Representation of the pre-synaptic terminal.} \textbf{A}: A Brownian particle (green) entering the pre-synaptic terminal either reaches a target (blue) located near the source points (orange) or reaches the bulk. In the bulk, the particle can bind to a substrate or leave the domain.
\textbf{B}: Schematic description of the arrival of a particle to a target site. The Brownian particles enter through source points. They can reach the small ribbon (red) located underneath a target (blue). The arrival of $T$ particles to the ribbon is needed to activate the target.}
\label{SchemaSpine}
\end{figure}

Cellular activation is described by the binding of few diffusing molecules to specific small and hidden targets. What is the time scale of a cell response, triggered by rare molecular events? Such a process requires exploring microdomains at a nanometer precision. Examples are vesicular fusion triggered by the binding of calcium ions, detection of a morphogen concentration by a growth cone, molecular events underlying the induction of synaptic plasticity at neuronal synapses or activating cellular check points where molecular events are transformed into a cellular decision.\\
Traditional chemical kinetics \cite{Redner} are based on mass action laws or reaction-diffusion equations, but there is a deficient description of stochastic chemical reactions in micro-domains, where only a small number of molecules are involved \cite{JCP2005,Burrage,Burrage2}. Several models were developed to describe stochastic reactions, such as Markov reaction-diffusion equations based on the joint probability distribution for the concentration of bound and unbound reactants, leading to coupled systems of reaction-diffusion \cite{Springer2015}. Another direction involves the Narrow Escape Theory \cite{Springer2015} to coarse-grain the binding and unbinding reactions to the time scale of diffusing reactant into and out of a binding site. This approach reduces the classical reaction-diffusion approach to a Markovian jump process description of the stochastic dynamics for the binding and unbinding of molecules. The Markovian approximation \cite{Redner,Daoduc} can be used for example to obtain predictions for the rate of molecular dynamics underlying spindle assembly checkpoint during cell division \cite{Daoduc2012} or the probability that a messenger RNA escapes degradation through binding a certain number of microRNAs. The Markovian approach can also be used to compute the mean time the number of bound molecules reaches a threshold, called the mean time to threshold theory (MTT), that characterizes the stability of activation \cite{Daoduc}. \\
However, none of the approaches described above can be used to model the interaction between a continuous bath of particles and discrete events where few particles bind to small targets for triggering a response. The goal of this letter is to introduce such a novel model and to provide an application from neurobiology for computing the probability of vesicle release following a calcium influx. Due to instrumental limitations, it is not yet possible to measure calcium dynamics at a nanometer level at neuronal synapses. Yet, synchronous and asynchronous synaptic release might depend on calcium dynamics and we provide here novel predictions by developing a novel hybrid stochastic model of chemical reactions at small and large scale, that we compare for confirmation to Gillespie simulations: the propose mechanism of asynchronous vesicular release clarifies also the role of calcium ions in short-term synaptic plasticity \cite{Nicoll} \\
{\noindent \bf Hybrid Markov-mass action model.} Reactant molecules are represented by Brownian particles, entering a domain $\Omega$ through localized point sources with an inward flux $J(t)$ through a single source point. These particles can reach independent target sites to trigger their activation. Following entrance, there are two possible fates for a particle: it can either reach the bulk where it is lost in the undifferentiated state of many particles, or it can hit a small target site (Fig.~\ref{MarkovModel}). The arrival of $T$ particles at a single target triggers activation. The Brownian reactant particles and the small targets can be described by a coarse-grained model, coupling the mass-action description (continuum level), while the discrete binding to the targets is described by a finite state Markov chain (Fig.~\ref{MarkovModel}).\\
For the application of vesicular release, the bulk represents the pre-synaptic terminal (Fig.~\ref{SchemaSpine}\textbf{A}), and the $N_{T}$ small target sites are hidden underneath a ball representing a vesicle (red ribbon in Fig.~\ref{SchemaSpine}\textbf{B}). The mean flux of particles entering the domain and reaching a target site $i$, for a stationary initial distribution $f(x)$ of source points is obtained by averaging over the distribution of vesicle locations
\beq
J^i(t) = <J^i(x,t)>_x = \int_{S} J(t)p_s(x)q(x,i)f(x)dx,
\label{MeanFlux}
\eeq
where $q(x,i)$ is the probability that a particle entering at point $x$ reaches the target site $i$.  The splitting probability $p_s$ that a particle hits a small target before entering the bulk has been estimated previously using conformal mapping methods and asymptotic approximations \cite{guerrierMFPT}:
 \beq
 p_s(d) = 1- \frac{1-A\ds\frac{r^2\eps}{H^3}}{1-\ds \frac{2 r \eps}{H^2}}\left(1-\frac{2 r\eps}{d^2} \right),
 \label{eqsplit}
\eeq
where $r$ is the size of the ball, $H$ is the distance to the bulk, $A=9.8$, $d$ is the distance to the closest target site, and $\eps$ is the height of the ribbon representing the site. Finally, The arrival rate of a Brownian particle coming from the bulk to a small ribbon (colored in red in fig. \ref{SchemaSpine}B) is Poissonian with rate \cite{guerrierMFPT}
\beq \label{lastf_Res}
k_T = \frac{4 \pi D \eps}{|\Omega|},
\eeq
where $D$ is the particle diffusion coefficient.\\
For each target site $i$ and for a given distribution ${\x} = (x^1,...,x^{N_p})$ of $N_p$ source points, the probability ${Pr}^i_{\x}\{k, t\}$ that $k$ particles are bound at time $t$ is computed from a Markov model: once a particle binds to a target, it cannot unbind, and thus the transition probability from the states $k-1$ to $k$ (bound particles), between $t$ and $t+\Delta t$, is due to the fluxes of particles from the point sources and from the bulk:
\beqq
{Pr}_{\x}^i\{k, t+\Delta t, {\x}\} = {Pr}_{\x}^i\{k-1, t\} \lambda^i_{dis}(t) \Delta t \\
+ {Pr}_{\x}^i\{k, t\}\left(1-\lambda^i_{dis}(t) \Delta t\right),
\eeqq
where $\lambda^i_{dis}(t)=\left(\sum_{l=1}^{N_p}J^i(x^l,t)+k_{T} N_f(t)\right)$, $k_{T}$ is the rate of arrival and $N_f(t)$ the number of free particles at time $t$ in the bulk.
A target is activated when there are exactly $T$ bound particles. We obtain for each target $1\leq i \leq N_{T}$, the Markov chain for the probabilities $ p_k^{\x,i}(t)={Pr}_{\x}^i\{k, t\},$
\beq\label{markov}
\frac{d}{dt}p_0^{\x,i}(t)&=&-\lambda^i_{dis}(t)p_0^{\x,i}(t) \nonumber \\
\frac{dp_k^{\x,i}(t)}{dt}&=&\lambda^i_{dis}(t)\left(p_{k-1}^{\x,i}(t)-p_k^{\x,i}(t)\right) \\
\frac{dp_T^{\x,i}(t)}{dt}&=&\lambda^i_{dis}(t)p_{T-1}^{\x,i}(t)\nonumber,
\eeq
where the last equation for $k=T$ describes the absorbing state for the probability $p_T^{\x,i}$. The initial condition at time $t=0$ is
\beq
p_k^{\x,i}(0) = \delta_{k=0} \hbox{ for } i=1..N_{T}
\label{InitCond}
\eeq
and the normalization condition is
\beq
\sum_{k=0}^{T} p_k^{\x,i}(t)=1 \hbox{ for } i=1..N_{T}.
\eeq
The time $\tau_T^i$ to threshold $T$ at target $i$ is defined as the mean first time that $T$ particles are located at target $i$. It is also the last passage time of a particle when $T-1$ particles are already at the binding sites. The distribution of time $\tau_T^i$ is called the activation time for the target $i$ and is given by
\beq
Pr\{\tau_T^i < t| {\x}\}=p_T^{\x,i}(t),
\label{tauTi}
\eeq
hence the probability density function $f_{\tau_T^i,{\x}}$ of activation times is
\beq \label{EqReleaseTime}
f_{\tau_T^i, {\x}}(t) = \frac{d p_T^{\x,i}(t)}{dt} = \lambda^i_{dis}(t) p_{T-1}^{\x,i}(t),
\eeq
and the mean release time $\bar \tau_T^i$ is thus
\beq
\bar \tau_T^i =\int_0^\infty t f_{\tau_T^i, {\x}}(t)dt.
\eeq
To conclude, we have presented here a Markov chain that counts for $T$ binding events and the particles originate either from the bulk or from a transient flux. We shall now study the coupling between a Markov chain and a density of free particles $N_f(t)$ that can also interact with a varying population of substrates.\\
{\noindent \bf The mass action law equations for reactant particles interacting in the bulk with a substrate.}
The reactant particles in the bulk can interact with a substrate according to the chemical reaction
\beq
S+R\xrightleftharpoons[\mathbf{k_{-1}}]{\mathbf{k_{0}}} S-R,
\eeq
where $k_{-1}$ (resp. $k_0$) is the backward (resp. forward) rate. We determine the mass action equations for the number of free $N_f(t)$, and bound $N_b(t)$ particles. The total number of substrate sites $S_{tot}$ is fixed, and at time $t$, the number of available sites is $S_{tot}-N_b(t)$. The particles can escape the domain $\Omega$ at a Poissonnian rate $k_{es}$, and bind to a target site with a rate $k_{T}$. We note that the probability that target $i$ is free, i.e. not activated, is given by
\beq
\sum_{k=0}^{T-1}p_k^{\x,i}(t) = 1-p_T^{\x,i}\left(t\right).
\eeq
In summary, $N_f$ and $N_b$ satisfy the mass action equations:
\beq\label{diff}
\frac{d N_f}{dt} &=& k_{-1} N_b - k_0 (S_{tot}-N_b)N_f + \sum_{l=1}^{N_p}\left(1- p_s(x^l) \right) J(t) \nonumber\\
&& - \left(k_{es} + k_{T}\sum_{i=1}^{N_{T}}\left(1- p_T^{\x,i}(t)\right) \right)N_f \\
&+&  T \sum_{i=1}^{N_{T}}\lambda_{dis}(t)p_{T-1}^{\x,i}(t), \nonumber \\
\frac{d N_b}{dt} &=&-k_{-1} N_b + k_0 (S_{tot}-N_b) N_f. \nonumber
\eeq
The first two terms in the first equations represent the classical unbinding and binding of molecules to buffers. The third term represents the fraction of particles directly entering the bulk from the initial influx. The fourth term corresponds to particles that leave the bulk or bind to a free target (we assume that the escape rate $k_{es}$ is fast compared to the binding rates). Finally the last term accounts for the release into the bulk of the $T$ particles bound to a target, following the target activation. Indeed, after activation, the $T$ bound particles are released into the bulk, leading to an increase of the free particles $N_f$ by a jump event of $T$ particles (eq.~\ref{EqReleaseTime}). The second equation is the classical mass action law for the number of bound buffers. \\
It is surprising that the system of equations \ref{markov}-\ref{diff}, for $t \geq t_0$ can be decoupled and resolved so we get 
\beq \label{proba}
p_k^{\x,i}(t) &=& \frac{1}{k!} \left(\int_{t_0}^t \lambda^i_{dis}(u) du\right)^k \exp\left(- \int_{t_0}^t \lambda^i_{dis}(u) du\right), \nonumber\\ &&\\
p_T^{\x,i}(t) &=&\exp\left(- \int_{t_0}^t \lambda^i_{dis}(u) du\right) \sum_{k\geq T}\frac{1}{k !}\left(\int_{t_0}^t \lambda^i_{dis}(u) du\right)^k. \nonumber
\eeq
To conclude, equations \ref{markov}-\ref{diff} involve the coupling of the probability $p^{\x,i}_T(t)$ to the bulk equation and the ensemble constitutes a coupled set of Markov-mass action equations that describes the cumulative binding of $T$ particles, interpreted as activation. We shall now provide an application in neurobiology about signaling at single neuronal synapses. There are indeed $10^{11}$ neurons in a human brain, and each has a mean of $10^3$ synapses. They are thought to code part of the memory, mainly by modulating the neuronal response through time variations in the release of vesicles. We will study now some aspect of the stochastic release of a vesicle from the accumulation of $T=5$ calcium ions to key proteins. This example illustrates the theory described above in the context of synaptic activation from rare molecular binding events.\\
{\bf Dynamics of vesicular release at neuronal synapse.}
The pre-synaptic terminal of a neuronal cell contains a large amount of vesicles that can be released with a certain probability following an action potential \cite{Rizzoli}. However, the release mechanism is complex and computing the release probability remains a challenge \cite{Meinrenken03,Dittrich,Matveev}. Following an action potential that invades the pre-synaptic terminal, voltage-gated calcium channels (VGCCs) open, leading to a calcium influx into the pre-synaptic domain. When several diffusing calcium ions have succeeded to find small molecular targets, which form a complex of molecular machinery (such as synaptotagmin and any other key molecules) located underneath a vesicle, these molecules are changing their conformation,  activating the fusion of the vesicle with the cell membrane and thus triggering the release of neurotransmitters \cite{Rizzoli}. This is a key step of neuronal communication. \\
At the same time, calcium ions can bind and unbind to buffer molecules located in the bulk of the pre-synaptic terminal. They can also exit at the end of the terminal, or be extruded by binding to small pumps, although this process is not completely documented. The success of vesicular release depends on the rare event triggered by the arrival of diffusing calcium ions to the small target molecules \cite{Neher,Neher2,Keller15}. This process also depends on the relative position between vesicles and the VGCCs and on their organization on the surface \cite{guerrierMFPT}.\\
 \begin{figure}[http!]
  \centering
  \includegraphics[scale=0.55]{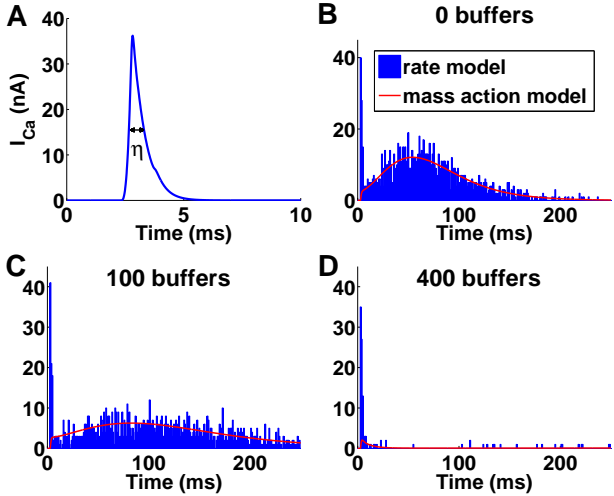}
\caption{ {\bf Calcium time course in the pre-synaptic terminal and vesicular release activation}. Channels are uniformly distributed over the membrane. \textbf{A}: Calcium current entry, simulated by a classical Hodgkin-Huxley model \cite{HilleBook}.
\textbf{B}-\textbf{D}: Histogram of vesicular release time for the stochastic (blue) and the Markov-mass action model (red) when there are zero (\textbf{B}), 100 (\textbf{C}), and 400 buffer sites (\textbf{D}). The continuous red curves are solutions of the hybrid model \ref{markov}-\ref{diff}, compared to the Poissonian simulations using the rate model.}
\label{CompSimu}
\end{figure}
To investigate the distribution time of released vesicles triggered by an accumulation of $T=5$ calcium ions at a ribbon (Fig~\ref{SchemaSpine}\textbf{B}), we consider one type of buffer in the synaptic bulk at various concentrations (Fig.~\ref{CompSimu}). We use equations \ref{diff}, when channels are uniformly distributed over the pre-synaptic terminal (we averaged the binding probabilities \ref{proba} over a uniform distribution). The initial influx of calcium ions through a VGCC is simulated using a simplified Hodgkin-Huxley model for calcium current \cite{HilleBook}. Indeed, the membrane depolarizes following an action potential, modeled by a transient applied current $I_{app}$: the calcium current transient (Fig.~\ref{CompSimu}\textbf{A}) lasts 2.75 ms, with a maximal value of $I_{Ca,\max}=36.2$ nA. The total charge $Q=0.025$ fC corresponds to an entry of 80 calcium ions. Fig.~\ref{CompSimu}\textbf{B} reveals that the distribution of vesicular release time is bimodal with two consecutive stages: a short time period that follows the immediate entrance of calcium ions, where vesicular release is triggered by ions that are directly coming from the channel influx. The second regime is characterized by a broader peak, induced by the random arrival of calcium ions from the bulk to the vesicular target sites, until $T$ sites per vesicle are occupied. The spread of the release time distribution is due to rare events of the diffusing calcium ions to the small targets.\\
To further clarify the origin of the residual (second) peak, we tested numerically the effect of changing the buffer concentration. Their role is to buffer the calcium in the bulk. We found that increasing the number of buffers abolishes the second peak (Fig.~\ref{CompSimu}\textbf{B}-\textbf{D}), confirming that it is due to the arrival of free calcium ions located in the bulk. We conclude with the following prediction: the diffusion time course in the bulk (which is a modeled as a ball of radius $600$ nm), where target binding sites are located underneath vesicles of size $20$ nm leads to a residual vesicular release that peaks around 60 ms. However, because the mean binding time to buffer is $1/k_{-1}=2s$, at the time scale of tens of ms, most ions are still buffered and the second peak disappears (Fig.~\ref{CompSimu}\textbf{C}-\textbf{D}) (the escape rate is $1/k_{es}=0.7s$).  However, we can still observe rare events in the distribution tail due to the arrival of free ions. To confirm this novel Markov-mass action model (red curves in Fig.~\ref{CompSimu}), we run also stochastic simulations using rate constants described above in a Gillespie algorithm (blue). We found that the two models quite different in nature agree together.\\
{\noindent \bf Discussion and conclusion.}
We presented here a Markov-mass action model to study the coupling between a continuum ensemble of molecules involved into a discreet number of rare events, representing the binding of a small and finite tiny subset of these Brownian molecules to small targets. It was already noticed that discrete-state stochastic models for particles are not well approximated by continuum equations \cite{Modchang,Weinberg2}. The present model is also quite different from previous numerical simulation efforts to generate a finite number of Brownian trajectories from a continuum ensemble \cite{nadler,Flegg}.\\
To illustrate the applicability of this method, we model vesicular release at synapses from calcium dynamics, where we found a first peak at short time scale (less than 10 ms) induced by the direct entrance of calcium ions, followed by a second peak, which is due to the return of calcium from the bulk to the small hidden targets, a process that also depends on the interaction with buffer molecules and can last hundreds of milliseconds. The emergence of these two phases can also explain asynchronous quantal release, without introducing any additional time constant at a molecular level. The present approach could also be used to describe short-term plasticity at a molecular level \cite{Keller15,Pan09} and to model the noise inherent to synaptic dynamics \cite{Peskin}.\\
{\bf Acknowledgement:} DH is currently a visiting Professor at the Mathematical Institute, Oxford OX2 6GG, UK. This research is supported by a Marie Curie Award and by the Simons foundation. D. H thanks the Newton Institute for its hospitality.


\end{document}